\documentclass[12pt]{iopart}
\usepackage{amsfonts}
\usepackage{amssymb}
\usepackage{graphicx}
\usepackage{dcolumn}
\usepackage{bm}
\usepackage{ccaption}
\usepackage{iopams}
\usepackage{setstack}
\usepackage{color}

\begin{document}
\bibliographystyle{unsrt}

\title{Coherent and Squeezed Vacuum Light Interferometry:\\Parity detection Hits the Heisenberg Limit}
\author{Kaushik P.\ Seshadreesan, Petr M.\ Anisimov, Hwang Lee, Jonathan P.\ Dowling}
\address{Hearne Institute for Theoretical Physics and Department of Physics and Astronomy \\
Louisiana State University, Baton Rouge, LA 70803 }
\ead{ksesha1@lsu.edu}

\date{\today}

\begin{abstract}
The interference between coherent and squeezed vacuum light can produce path
entangled states with very high fidelities. We show that the phase sensitivity of the above interferometric scheme with parity
detection saturates the quantum Cramer-Rao bound, which reaches the Heisenberg-limit when the coherent and
squeezed vacuum light are mixed in roughly equal proportions. For the same interferometric scheme, we draw a detailed comparison between parity detection and a symmetric-logarithmic-derivative-based detection scheme suggested by Ono and Hofmann.
\end{abstract}


\maketitle

\section{Introduction}
Optical metrology relies on light interferometry as its primary tool for phase
estimation. The sensitivity of phase estimation with coherent light based interferometry is limited by shot noise~\cite{caves}. This limit,
however, is due to the classical nature of coherent light and can be surpassed if
nonclassical states of light, such as the N00N state, are
used~\cite{jonmetrology, barry}. Still there is a limit on the
sensitivity of phase estimation in the case of linear optical interferometry. Its usual justification stems from the Heisenberg uncertainty principle that links phase uncertainty of a state to its photon number uncertainty, $\Delta\phi\Delta n\geq
1$. Combination of this equation with
the assumption that the photon number uncertainty in a state is limited by the total photon
number (in the case of states with definite photon number) or the total average photon number (in the case of states with indefinite photon number), $\Delta n\le N$, suggests the limiting phase sensitivity to
be $\Delta\phi_{\rm HL}=1/N$, which is commonly referred to as the Heisenberg
limit~\cite{ou_2, holbur}. 

Quantum optical metrology has Heisenberg limited sensitivity of phase
estimation as its goal. To this end, search for convenient states of light and
optimal detection schemes still continues~\cite{dorner, uys, obrien, nagata, higgins, pezzi_2}. Candidate states of light are gauged based on the quantum Cramer-Rao bound that provides a detection scheme independent phase sensitivity $\Delta \phi_{\rm QCRB}$~\cite{QCRB}. In turn, optimal detection schemes are sought, which are capable of saturating the quantum Cramer-Rao bound. A known possibility is a detection scheme that measures a symmetric logarithmic derivative, since such operators saturate the quantum Cramer-Rao bound; however, they are seldom easy to implement. The capabilities of alternative detection schemes are judged by the
classical Cramer-Rao bound that is detection scheme specific, or by the error
propagation formula that links the uncertainty of the
observed signal with phase uncertainty.

\begin{figure}[h]\centering
\includegraphics[scale=1.0]{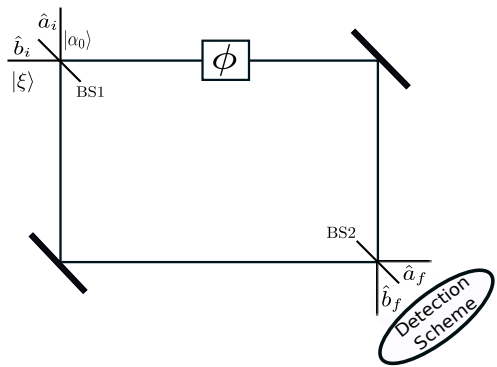}
\caption{A lossless Mach-Zehnder Interferometer with two-mode state input given by the product of coherent and squeezed vacuum states of light. The modes are labeled by the annihilation operators: $\hat{a}_{i}$, $\hat{b}_{i}$ for the input ports and $\hat{a}_{f}$, $\hat{b}_{f}$ for the final output ports, respectively. BS1 and BS2 are 50-50 beam splitters.}
\label{fig:CohSV}
\end{figure}

Here, we consider coherent and squeezed vacuum light input as a candidate for Heisenberg limited phase estimation (see Fig.~\ \ref{fig:CohSV}). This state has been previously
checked against the quantum Cramer-Rao bound and shown to achieve Heisenberg limited phase sensitivity when the coherent and squeezed vacuum light are mixed in roughly equal intensities~\cite{pezzismerzi_1}---a feature that can be explained as due to the high fraction of a N00N state in the normalized N-photon output component of the quantum state of light past the mixing beam splitter---first pointed out by Hofmann and Ono~\cite{hofmannono} and later experimentally demonstrated by Silberberg's group~\cite{silberberg}. The detection scheme suggested in Ref.~\cite{pezzismerzi_1} for Heisenberg limited phase estimation was however based on Bayesian analysis of the photon number statistics of the output state, which requires photon number counting in both modes.

In this paper, we study parity detection~\cite{gerry2} for the interferometry with coherent and squeezed vacuum light. We show that the parity operator saturates the quantum Cramer-Rao bound, and in turn provides Heisenberg limited phase sensitivity when the coherent and squeezed vacuum light are mixed in equal proportions. Parity detection should be a simpler alternative to the detection scheme of Ref.~\cite{pezzismerzi_1} since parity measurement can be inferred from the photon number counting statistics of a single mode alone, which in the low power regime can be obtained using photon-number-resolving detectors~\cite{migdall}. Although accurate photon-number-resolution in the high power regime is a difficult task, there have been proposals for the quantum non-demolition measurement of photon number using weak non-linearities and homodyning~\cite{nonlinear2}. On the other hand, it is not necessary to have photon-number-resolving capabilities in order to implement parity detection. Assuming the availability of large Kerr nonlinearities through the techniques of electromagnetically induced transparency, a scheme that performs quantum non-demolition measurement of parity directly, without requiring the measurement of photon number, has been proposed~\cite{nonlinear}. Plick et {\it al.} have recently shown that parity measurement for interferometric schemes that use Gaussian states, like the one in use here, could possibly be inferred through balanced homodyning and intensity difference measurement~\cite{bill}.

Ono and Hofmann, in Ref.~\cite{onoscheme}, have studied a detection scheme based on the measurement of a symmetric logarithmic derivative for the considered interferometric scheme. This scheme uses interference with an auxiliary local oscillator and intensity difference measurement as well. We dutifully discuss this scheme with an intent to compare it with parity detection.

The paper is organized as follows. Section~\ref{machinery} describes the propagation of a two mode light, initially in the product state of coherent
and squeezed vacuum light, through the Mach-Zehnder interferometer. Section~\ref{paritymeasurement} focuses on the parity-based detection scheme and
provides the expected signal and phase sensitivity, while section~\ref{disc} discusses the Ono-Hofmann detection scheme in equal detail. Section~\ref{concl} deals with the conclusion. 

\section{Propagation of the input fields through the Interferometer}
\label{machinery}
The input to the interferometer is in the product state 
$|\alpha_{0}\rangle\otimes|\xi=r\ e^{i\phi_s}\rangle$ that describes
coherent light with displacement $\alpha_{0}=\sqrt{n_{c}}e^{-i\phi_{c}}$ in one mode and squeezed vacuum with parameters $r$ and $\phi_s$ in the other. The corresponding Wigner function of the input state is the product of the respective Wigner functions as
well~\cite{gerryknight}:
\begin{equation}
W_{\rm in}(\alpha, \alpha_0; \beta, r)=W_{c}(\alpha,\alpha_0) W_{s}(\beta,r),
\end{equation}
with Wigner function for the corresponding states being
\begin{equation}
\begin{array}{c}
\label{WcWs}
W_{c}(\alpha,\alpha_0)=\frac{2}{\pi}e^{-2|\alpha-\alpha_0|^2},\ 
W_{s}(\beta,r)=\frac{2}{\pi}e^{-2|\beta|^2 \cosh2r-\left(\beta^2 +\beta^{*2} \right)\sinh2r },
\end{array}
\end{equation}
and where we have made $\phi_s=0$ by appropriately fixing the irrelevant absolute
phase. This choice implies that the phase of the coherent light $\phi_{c}$ is now
measured with respect to the phase of the squeezed vacuum state.

A Mach-Zehnder interferometer is composed of optical elements such as beam
splitters, mirrors and phase shifters. Propagation of the light field
through these elements is described by relating the initial variables in the
Wigner function to their final expressions: 
\begin{equation}
W_{\rm out}(\alpha_f,\beta_f)=W_{\rm in}(\alpha_i(\alpha_f,\beta_f),\beta_i(\alpha_f,\beta_f)).
\label{eq:WafterBS}
\end{equation}
The relation between variables in the most general form is given by a two-by-two scattering matrix $\hat{M}$: 
\begin{equation}
\left[
\begin{array}{c}
\alpha_i\\
\beta_i
\end{array}
\right]=\hat{M}^{-1}
\left[
\begin{array}{c}
\alpha_f\\
\beta_f
\end{array}
\right],
\label{eq:AMPTRANS}
\end{equation}
where $\alpha_i$, $\beta_i$, $\alpha_f$, and $\beta_f$ represent the complex
amplitudes of the field in the modes $\hat{a}_i$, $\hat{b}_i$, $\hat{a}_f$,
and $\hat{b}_f$, respectively. More specifically, propagation through a 50-50 beam-splitter and a phase shifter (in mode $\hat{b}$), are described
by:
\begin{eqnarray}
\hat{M}_{\rm BS}=\frac{1}{\sqrt{2}}  \left[
\begin{array}{lr}
1   & i \\
i & 1
\end{array}
\right] ,\
\label{eq:UxBS}
\hat{M}_\phi=\left[
\begin{array}{lr}
1    & 0 \\
0 & e^{-i\phi}
\end{array}
\right],
\label{eq:UxPh}
\end{eqnarray}
respectively. Therefore, the Mach-Zehnder interferometer in
Fig.~\ref{fig:CohSV} is described by $\hat{M}_{\rm
  MZI}=\hat{M}_{\rm BS}\hat{M}_\phi\hat{M}_{\rm BS}$ and is found to be:
\begin{equation}
\hat{M}_{\rm MZI}=
i e^{-i \frac{\phi}{2}}\left[
\begin{array}{cc}
\sin\frac{\phi}{2} & \cos\frac{\phi}{2}\\
 \cos\frac{\phi}{2}& -\sin\frac{\phi}{2}
\end{array}
\right],
\end{equation}
with the corresponding transformation of the variables in the following form:
\begin{eqnarray}
\alpha_{i} \rightarrow -i e^{i \frac{\phi}{2}} (\alpha _f\sin\frac{\phi}{2} +  \beta _f\cos\frac{\phi}{2}),\\
\beta_{i} \rightarrow  -i e^{i \frac{\phi}{2}} (\alpha _f\cos\frac{\phi}{2}-\beta _f\sin\frac{\phi}{2}).
\end{eqnarray}
Therefore, the state of light at the output of the Mach-Zehnder interferometer
is described by the following Wigner function:
\begin{eqnarray}
\label{Wout}
W_{\rm out}(\alpha_f, \beta_f)&=&\frac{4}{\pi^2}e^{-2|i e^{i \frac{\phi}{2}} (\alpha _f\sin\frac{\phi}{2}+\beta _f\cos\frac{\phi}{2})+\alpha_0|^2}\nonumber\\
&\times& e^{ -2|\alpha_f\cos \frac{\phi }{2}-\beta_f\sin\frac{\phi }{2}|^2\cosh2r}\nonumber \times e^{2\ {\rm Re}\left[e^{i\phi }\left(\alpha_f\cos \frac{\phi }{2}-\beta_f\sin \frac{\phi }{2}\right)^2\right]\sinh 2r}.
\end{eqnarray}

Having found the state of light at the output of the Mach-Zehnder interferometer, we will present the parity-based phase estimation scheme with calculations of its signal and phase sensitivity in the following section.

\section{Phase Estimation with Parity Measurement}
\label{paritymeasurement}

Parity detection was originally proposed in the context of trapped ions by
Bollinger \emph{et al.}~\cite{bollinger}. It was later adopted for optical
interferometry by Gerry~\cite{gerry}. In its essence, parity detection
distinguishes states with odd and even number of photons. Quantum mechanically,
it is described by the parity operator,
$\hat{\Pi}_{a}=(-1)^{\hat{a}^{\dagger}\hat{a}}$, acting on a single output mode, $\hat{a}$. Parity detection makes phase inference at the Heisenberg limit possible without
having to know the full photon number counting statistics for several classes of input states with definite as well as indefinite photon
numbers (including the N00N state)~\cite{aravindlee, petr}. Coherent and squeezed vacuum light belong to the latter
class of states and the performance of parity detection for these states is
studied in this section. Although the fact that parity detection achieves Heisenberg limited performance with the N00N-state is a motivation for this study, it is by no means a reason in itself to believe that parity detection would work equally well for the considered interferometric scheme. This is because the quantum state in the interferometer also has several miscellaneous contributions apart from that of a N00N state.

An expected signal of the parity detection scheme
$\langle\hat{\Pi}_{a}\rangle$ is calculated as the value of the Wigner
function at the origin for the corresponding mode. In the case of mode
$\hat{a}_f$, $\langle\hat{\Pi}_{a_{f}}\rangle=\frac{\pi}{2}\int W_{\rm out}(0,\beta)
d^{2}\beta$, and is found to be:
\begin{equation}
\langle\hat{\Pi}_{a_{f}}\rangle=\frac{\exp\left[-n_{c} \left(\frac{\sqrt{n_{s}^2+n_{s}} \sin ^2\phi \cos 2  \phi_{c} -\cos \phi }{n_{s} \sin ^2 \phi +1}+1\right)\right]}{\sqrt{n_{s} \sin ^2 \phi +1}},
\label{eq:sparity}
\end{equation}
where the coherent light amplitude and the squeezing parameter have been expressed in
terms of the average photon numbers, $n_{c}$ and $n_{s}$, using the relations
$\alpha_{0}= \sqrt{n_{c}} e^{-i \phi_{c}}$ and $r=\sinh^{-1} \sqrt{n_{s}}$.

\begin{figure}[h]\centering
\includegraphics[scale=0.8]{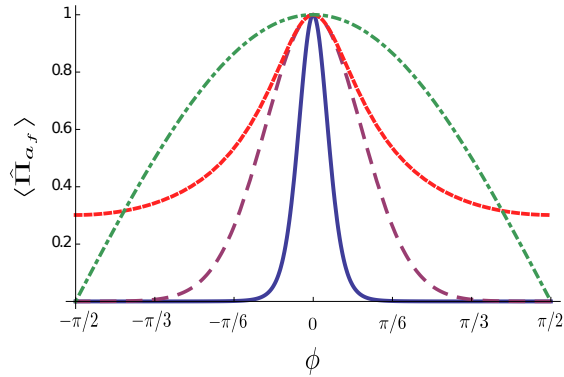}
\caption{(Color online) The parity signal, $\langle\hat{\Pi}_{a_{f}}\rangle$, as a function of the accumulated phase difference between the arms of the MZI, $\phi$: dashed (magenta) line for coherent light interferometry ($\eta=0$) with $n_{c}=10$; dotted (red) line for squeezed vacuum light interferometry ($\eta=1$) with $n_s=10$; and solid (blue) line for coherent and squeezed vacuum light interferometry ($\eta=0.5$) with $n_{c}=n_{s}=5, \ \phi_c=0$. The dot-dashed (green) line is the signal for conventional coherent light interferometry with intensity difference measurement.}
\label{paritysignal}
\end{figure}

The signal of the parity detection scheme is periodic with period $2\pi$ and
attains its maximum value of one at $\phi=0$. Although this maximum value is
independent of the phase of the coherent
light $\phi_{c}$ and the light intensities $n_{c}$ and $n_{s}$, the visibility of the signal and its width are functions of
these parameters. The visibility of the signal is found to be best when $\phi_c=0$ and to diminish as $\phi_c$ drifts away from zero, becoming worst at $\phi_c=\pi/2$. Since it is reasonable to assume the coherent and squeezed vacuum light to be locked to the same external phase, $\phi_c$ can be set to zero for optimal performance. Further, the dependence of the signal on the light intensities is studied in terms of the total input intensity, 
$n_{\rm in}=n_{c}+n_{s}$, and the fraction of total intensity in the squeezed vacuum
state, $\eta=n_{s}/n_{in}$. When $\eta$ is increased from zero, the signal is found to grow narrower until reaching an optimal width, and then to broaden again, but with reduced visibility as $\eta$ approaches one. For $\eta=0$ and
$\eta=1$, the width of the signal is found to be proportional to $\pi/\sqrt{n_{\rm in}}$, which is
narrower than the resolution of conventional interferometry by a factor of $\sqrt{n_{\rm in}}$ and thus demonstrates super-resolution~\cite{superresolution}. The fraction $\eta=0.5$ is found to be the most optimal choice for distributing the input light intensity, since it allows a higher narrowing
factor of $n_{\rm in}$. Figure~\ref{paritysignal} demonstrates this result by
comparing the parity signals for interferometry with only coherent light
($\eta=0$)~\cite{gao} or squeezed light ($\eta=1$) and interferometry with coherent and
squeezed vacuum light of equal intensities ($\eta=0.5$). We see that for the same total input photon number, $n_{\rm in}=10$, the parity
signal for the latter case is narrower than any other case.

The phase sensitivity $\Delta \phi$ of an interferometer, followed by a
detection  scheme described by an operator $\hat{O}$, can be characterized
using the error propagation formula:
\begin{equation}
\label{error}
\Delta\phi^{2}=\frac{\langle
  \hat{O}^{2}\rangle-\langle\hat{O}\rangle^2}{\left| d \langle \hat{O}\rangle/d\phi\right|^2}.
\end{equation}
For the parity based detection scheme, $\hat{O}~=~\hat{\Pi}_{a_{f}}$,
knowing the signal suffices for sensitivity calculation since
$\hat{\Pi}^{2}_{a_{f}}=1$. The phase sensitivity with parity detection for
coherent and squeezed vacuum light interferometry is found to be best at $\phi=0$, and is given by:
\begin{equation}
\Delta\phi^2=\frac{1}{2 n_{c}\sqrt{n_{s}(n_{s}+1)}\cos2\phi_c+2 n_{c}n_{s}+n_{c}+n_{s}}.
\label{parity_optimal}
\end{equation}

For a detection scheme to be optimal, it has to saturate the quantum
Cramer-Rao bound. The quantum
Cramer-Rao bound for the considered interferometric scheme was derived in Ref.~\cite{pezzismerzi_1} and found to be:
\begin{equation}
\Delta\phi_{\rm QCRB}^2=\frac{1}{|\alpha_{0}|^2e^{2r}+\sinh^2r}.
\label{QCRB}
\end{equation}
This expression can be shown to be identical to the phase sensitivity with parity detection, given
in Eq.~(\ref{parity_optimal}) (under the condition $\phi_{c}=0$), when $\alpha_{0}$ and $r$ are replaced in terms of the
average photon numbers, $n_c$ and $n_s$. Thus, parity detection saturates the quantum Cramer-Rao bound and is optimal for the
considered interferometric scheme.
 
\begin{figure}[h]\centering
\includegraphics[scale=0.8]{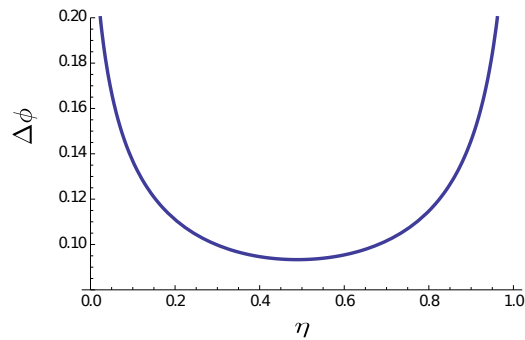}
\caption{(Color online) Phase sensitivity with parity detection for coherent and squeezed vacuum light interferometry $\Delta\phi$ as a function of the fraction of squeezed vacuum in the input $\eta$. The total input photon number $n_{\rm in}=10$ and $\phi_c=0$.}
 \label{QCRB_sens}
\end{figure}

\begin{figure}[h]\centering
\includegraphics[scale=0.8]{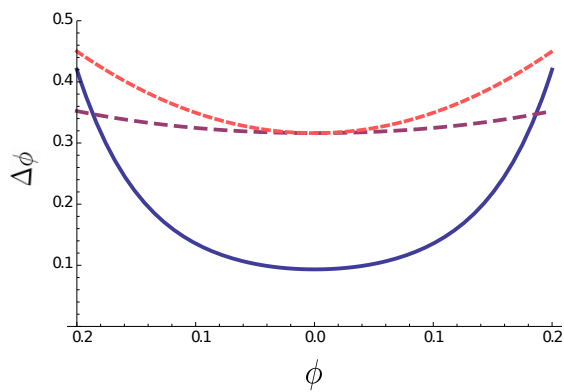}
\caption{(Color online) Phase sensitivity with parity detection $\Delta\phi$ as a function of the accumulated phase difference between the arms of the MZI $\phi$: dashed (magenta) line for coherent light interferometry ($\eta=0$) with $n_{c}=10$, $\phi_c=0$, dotted (red) line for squeezed vacuum interferometry ($\eta=1$) with $n_{s}=10$ and solid (blue) line for coherent and squeezed vacuum light interferometry ($\eta=0.5$) with $n_{c}=n_{s}=5$, $\phi_c=0$.}
 \label{paritysensitivity}
\end{figure}

\begin{figure}[h]\centering
\includegraphics[scale=0.8]{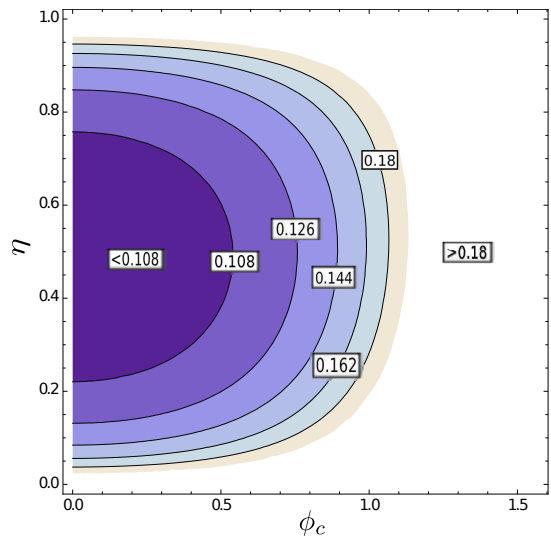}
\caption{(Color online) Contour plot of the phase sensitivity with parity detection for coherent and squeezed vacuum light interferometry close to phase origin, as a function of the phase of the coherent light $\phi_{c}$ and the fraction of squeezed vacuum in the input $\eta$, respectively. The total photon number $n_{\rm in}=10$. At the point $\eta=0.5$ and $\phi_c=0$, phase sensitivity attains it's optimum value of $0.093$.} 
\label{paritycontour}
\end{figure}

Although parity detection is optimal for the considered interferometric
scheme irrespective of the input intensities, the combination as a whole achieves its best phase sensitivity when
$\eta=0.5$. Figure~\ref{QCRB_sens} is a plot of the 
phase sensitivity $\Delta\phi$ given in Eq.~(\ref{parity_optimal}) (under the condition $\phi_{c}=0$), as a function of the fraction of squeezed vacuum in the input $\eta$. The phase sensitivity can be seen to be best when $\eta\approx 0.5$. Equation~(\ref{parity_optimal}), under the condition $\phi_{c}=0$, reveals that the phase sensitivity of the combination coincides with the Heisenberg-limit, $\Delta\phi\approx 1/n_{\rm in}$, when $\eta\approx 0.5$, while it coincides with the shot-noise limit, $\Delta\phi\approx\ 1/\sqrt{n_{\rm in}}$, when $\eta=0$ or 1. Figure~\ref{paritysensitivity} illustrates the contrast in phase sensitivities with parity detection for the cases corresponding to coherent and squeezed vacuum light interferometry with $\eta=0$, $\eta=1$, and $\eta=0.5$.

Finally, we study the critical dependence of phase sensitivity on the fraction of squeezed vacuum $\eta$ and the input phase $\phi_c$ near the optimal point of operation, namely $\eta=0.5,\ \phi_{c}=0$. The contour plot given in Fig.~\ref{paritycontour} shows that with small variations in $\eta$ and $\phi_c$, the phase sensitivity of the scheme is not much compromised. Thus, the parity detection scheme provides Heisenberg limited phase sensitivity even while allowing for small fluctuations in $\eta$ and $\phi_c$ about the optimal point of operation.

\section{Discussion}
\label{disc}

\begin{figure}[h]\centering
\includegraphics[scale=1]{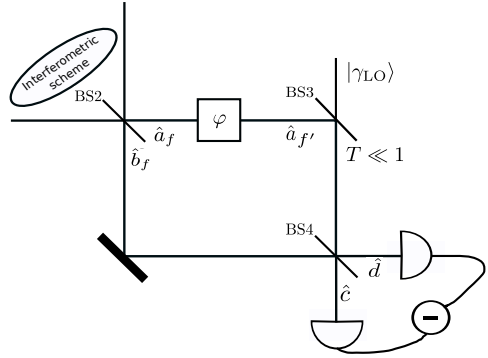}
\caption{The Ono-Hofmann detection scheme for interferometry with coherent and squeezed vacuum light. The detection scheme uses interference with an auxiliary local oscillator and intensity difference measurement for phase estimation. A highly reflective beam splitter is used to mix the local oscillator field into the interferometer.}
\label{fig:CohSVOno}
\end{figure}

So far, we have shown that parity detection could be used to achieve Heisenberg-limited phase estimation in the interferometry with coherent and squeezed vacuum light. The different possible implementations of parity detection include the use of photon-number-resolving detectors in the low power regime; Kerr nonlinearities and homodyning in the high power regime. In Ref.~\cite{onoscheme}, Ono and Hofmann discussed a different detection scheme that implements the measurement of a symmetric logarithmic derivative. Implementation of this measurement is based on interference with a local oscillator and intensity difference measurement (see Fig.~\ref{fig:CohSVOno}). Since symmetric logarithmic derivative based phase estimators saturate the quantum Cramer-Rao bound, Heisenberg-limited phase sensitivity was anticipated with this scheme for the interferometry with coherent and squeezed vacuum light mixed in equal proportions ($\eta=0.5$). Here, we present a brief study of the Ono-Hofmann detection scheme (in the absence of losses), for the purpose of comparing it with parity detection.

The Ono-Hofmann detection scheme consists of a second MZI appended at the output of the first, with a control phase $\varphi$, which is set to zero \footnote{It was set to $\pi$ by Ono and Hofmann in their original analysis. We choose zero for ease of calculation. The phase sensitivities in both cases turn out to be identical, although the signals differ.}. A local oscillator field, which is in the coherent state, $|\gamma_{\rm lo}\rangle~=~\sqrt{n_{\rm lo}/T}e^{i\phi_{\rm lo}}$, is introduced by mixing with the mode $\hat{a}_{f'}$ through a highly reflective beam splitter of transmissivity, $T<<1$, where $n_{\rm lo}$ is the average number of photons in the field that eventually enters the interferometer, and $\phi_{\rm lo}$, its phase. In the end, the difference in intensities at the two output modes is measured.

Intensity measurements at the output provide:
\begin{equation}
\begin{array}{lcl}
\langle\hat{c}^\dagger\hat{c}\rangle=\langle\left\{\hat{c}^\dagger\hat{c}\right\}_s\rangle-\frac{1}{2},\\
\langle\hat{d}^\dagger\hat{d}\rangle=\langle\left\{\hat{d}^\dagger\hat{d}\right\}_s\rangle-\frac{1}{2},
\end{array}
\end{equation}
$\left\{\hat{c}^\dagger\hat{c}\right\}_s$ ($\left\{\hat{d}^\dagger\hat{d}\right\}_s$) being the symmetric form of the operator, which can be evaluated based on the final Wigner function of the state $W_f$ as:
\begin{eqnarray}
\label{wigner_1}
\langle\left\{\hat{c}^\dagger\hat{c}\right\}_s\rangle=\int \int |\alpha|^2W_f(\alpha,\beta)d^2\alpha d^2\beta,\\
\langle\left\{\hat{d}^\dagger\hat{d}\right\}_s\rangle=\int \int |\beta|^2W_f(\alpha,\beta)d^2\alpha d^2\beta,
\end{eqnarray}
where $\alpha$ and $\beta$ are the complex amplitudes in the modes $\hat{c}$ and $\hat{d}$ respectively.

\begin{figure}[h]\centering
\includegraphics[scale=0.8]{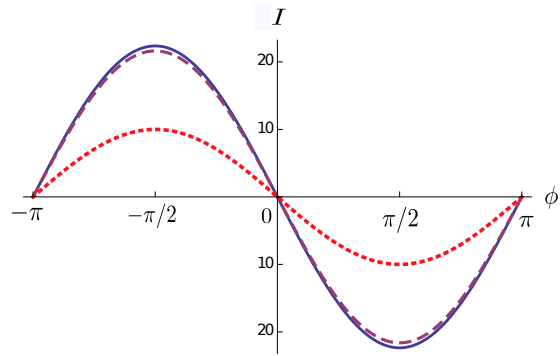}
\caption{(Color online) The signal with the Ono-Hofmann detection scheme---the intensity difference $I$ is plotted as a function of the accumulated phase between the arms of the MZI $\phi$: dashed (magenta) line for coherent light interferometry ($\eta=0$) with $n_{c}=10,\ \phi_c=0$, dotted (red) line for squeezed vacuum light interferometry ($\eta=1$) with $n_{s}=10$ and solid (blue) line for coherent and squeezed vacuum light interferometry ($\eta=0.5$) with $n_{c}=n_{s}=5,\ \phi_c=0$. A local oscillator of strength $n_{\rm lo}=100$ and phase $\phi_{\rm lo}=\pi/2$ is used.}
\label{fig:onosignal}
\end{figure}

The signal, which is the difference in intensities at the output ports, is thus given by:
\begin{equation}
\begin{array}{lc}
 I=\int \int \left(|\alpha|^2-|\beta|^2\right)W_f(\alpha,\beta)d^2\alpha d^2\beta,
\end{array}
\end{equation} 
and is found to be:
 \begin{equation}
 \begin{array}{lcl}
 I=-2 \sqrt{n_c n_{\rm lo}}\cos\frac{\phi }{2}\cos\left(\frac{\phi }{2}+\phi_{c} -\phi_{\rm lo} \right)+\left(n_c-n_{s}\right)\sin\phi.
 \end{array}
 \label{onosig}
 \end{equation}
It is plotted in Fig.~\ref{fig:onosignal}, as a function of $\phi$, under the condition $\phi_c=0,\ \phi_{\rm lo}=\pi/2$ (the point where the phase sensitivity is found to be optimal, as mentioned later in the paper). The figure compares the signal for the interferometry with equal intensities of coherent and squeezed vacuum light ($\eta=0.5$), with those of interferometry with only coherent light ($\eta=0$) and only squeezed vacuum light ($\eta=1$). We see that for the same total input photon number, $n_{\rm in}=10$, and same number of photons from the local oscillator, $n_{\rm lo}=100$, the signal for the former is stronger than any other case. However, unlike with parity detection, there is no super-resolution in the signal for the Ono-Hofmann detection scheme.

\begin{figure}[h]\centering
\includegraphics[scale=0.8]{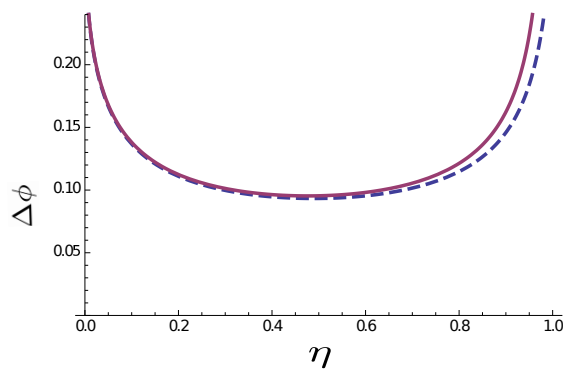}
\caption{(Color online) Phase sensitivity of coherent and squeezed vacuum light interferometry as a function of the fraction of squeezed vacuum light in the input $\eta$ $n_{\rm in}=10$: solid (magenta) line for phase sensitivity with the Ono-Hofmann detection scheme, $\Delta\phi$, dashed (blue) for the Quantum Cramer-Rao bound, $\Delta\phi_{QCRB}$.}
\label{ono_sens_2}
\end{figure}

Phase sensitivity with the Ono-Hofmann detection scheme for the interferometry with coherent and squeezed vacuum light is calculated based on the error propagation formula mentioned in Eq.~(\ref{error}). Variance of the signal $\Delta I^2$ which is required in the formula, can be shown to be:
\begin{equation}
\label{variance}
\begin{array}{ll}
\Delta I^2=
\int \int\left( |\alpha|^2- |\beta|^2\right)^2W_f(\alpha,\beta)d^2\alpha d^2\beta-\frac{1}{2}.
\end{array}
\end{equation}
The phase sensitivity thus calculated, is found to be optimal at $\phi=\pi$, under the condition $\phi_{c}=0,\ \phi_{\rm lo}=\pi/2$, and given by:
\begin{equation}
\label{ono1}
\begin{array}{ll}
\Delta\phi^2=
\frac{-2 \left(-2 n_{s}+2 \sqrt{n_{s}+1} \sqrt{n_{s}}-1\right) (\sqrt{n_{c}}-\sqrt{n_{\rm lo}})^2+2 n_{s}+1}{2 (\sqrt{n_{c}} (\sqrt{n_{\rm lo}}-\sqrt{n_{c}})+n_{s})^2}.
\end{array}
\end{equation}
In the limit of infinite $n_{\rm lo}$, it takes the following simplified form:
\begin{equation}
\label{ono_sens}
\begin{array}{lcl}
\Delta\phi^2=\frac{2 n_{s}-2 \sqrt{n_{s}^2+n_s}+1}{n_{c}}\approx \frac{1}{4n_c n_s} \ {\rm for}\ (n_s>1).\\
\end{array}
\end{equation}

\begin{figure}[h]\centering
\includegraphics[scale=0.8]{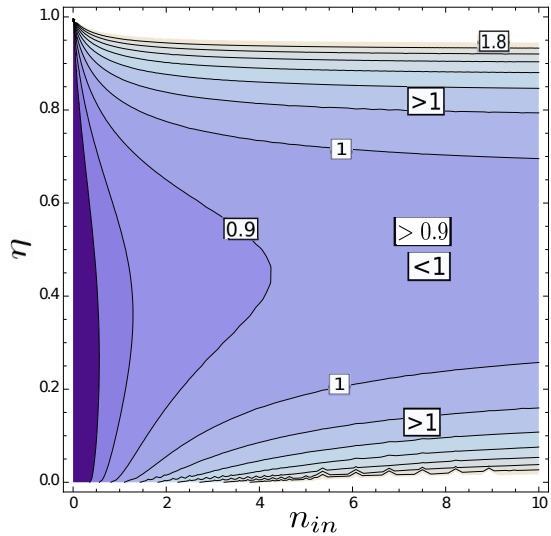}
\caption{(Color online) The Ono-Hofmann detection scheme for the interferometry with coherent and squeezed vacuum light: A contour plot of $\Delta\phi\times n_{\rm in}$ with respect to $n_{\rm in}$ and the fraction of squeezed vacuum light in the input $\eta$ in the presence of a local oscillator of infinite power. The contour of value one corresponds to points of Heisenberg-limited phase sensitivity. It can be seen that this contour tends to $\eta=0.5$ in the limit of large $n_{\rm in}$.} 
\label{ono_contour_1}
\end{figure}

\begin{figure}[h]\centering
\includegraphics[scale=0.8]{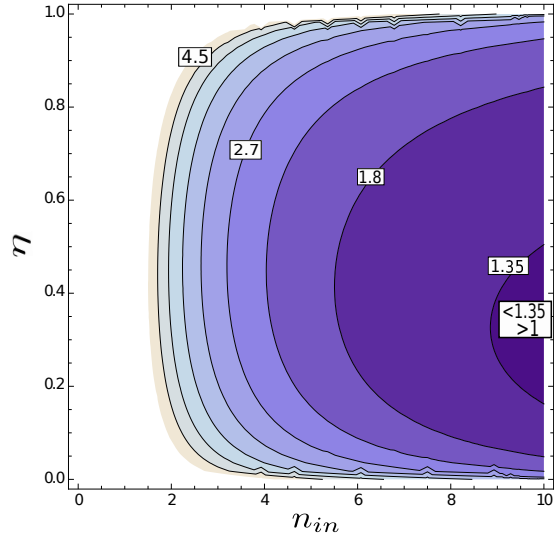}
\caption{(Color online) The Ono-Hofmann detection scheme for the interferometry with coherent and squeezed vacuum light: A contour plot of $\Delta\phi$ $\times \sqrt{n_{t}}$ with respect to $n_{\rm in}$ and the fraction of squeezed vacuum light in the input $\eta$ in the presence of a local oscillator with photon number $n_{\rm lo}=100$.  A contour of value one would correspond to points of shot noise-limited phase sensitivity, while contours of values above one refer to points where phase sensitivity is worse than shot noise.} 
\label{ono_contour_2}
\end{figure} 

This expression {\it nearly} saturates the quantum Cramer-Rao bound for the interferometry with coherent and squeezed vacuum light given in Eq.~(\ref{QCRB}), as can be seen in Fig.~\ref{ono_sens_2}, which compares graphs of the quantum Cramer Rao bound and the phase sensitivity given in Eq.~(\ref{ono_sens}), as a function of the fraction of total intensity in the squeezed vacuum state $\eta$ for a given total input intensity, $n_{\rm in}=10$. The fact that they don't completely overlap can be explained as due to the use of the error propagation formula instead of the more accurate classical Cramer Rao bound, which in this case is difficult to calculate.

Since we are mainly interested in achieving Heisenberg limited phase sensitivity with the considered interferometric scheme, we now focus our attention on the specific case corresponding to $\eta=0.5$, since the quantum Cramer Rao bound of Eq.~(\ref{QCRB}) for this $\eta$ matches the Heisenberg limit. Having found that the Ono-Hofmann detection scheme saturates the quantum Cramer Rao bound (in the limit of infinite $n_{\rm lo}$), we look to estimate the amount of local oscillator power that is required to provide Heisenberg-limited phase estimation, for a given total input photon number, $n$. In other words, we look to make an order of magnitude estimate for the infinity in terms of total input intensity. When $n_s\approx n_c \approx \frac{n_{\rm in}}{2}$, the expression in Eq.~(\ref{ono1}) takes the form: 
\begin{equation}
\label{ono2}
\begin{array}{lcl}
\Delta^2\phi=
\frac{\left(n_{\rm in}+1-\sqrt{n_{\rm in}^2+2n_{\rm in}}\right) \left(n_{\rm in}+2n_{\rm lo}-\sqrt{8n_{\rm in}n_{\rm lo}}\right)+n_{\rm in}+1}{n_{\rm in}n_{\rm lo}}.
\end{array}
\end{equation}
In the limit of large $n_{\rm in}$, the regime of interest of the Ono-Hofmann detection scheme, this can be expanded in a series as:
\begin{equation}
\Delta\phi=\frac{1}{\sqrt{n_{\rm lo}}}+\frac{3}{4n_{\rm in}\sqrt{n_{\rm lo}}}-\frac{1}{\sqrt{2 n_{\rm in}^3}}
\end{equation}
From the above expansion, it is evident that the local oscillator photon number, $n_{\rm lo}$, has to be as large as $n_{\rm in}^2$ for the scheme to provide Heisenberg-limited phase sensitivity. Thus, if one were to include the local oscillator in the photon budget, the phase sensitivity that the Ono-Hofmann detection scheme can be expected to provide is at best shot-noise limited. This shortcoming is illustrated using Figs.~\ref{ono_contour_1} and \ref{ono_contour_2}, which show contour plots drawn with respect to the total input photon number $n_{\rm in}$ and the fraction of total input intensity in the squeezed vacuum state $\eta$. Although the total photon number $n_{\rm in}$ in the plots run only up to ten, they still illustrate the point. Figure \ref{ono_contour_1} shows the plot of the product of phase sensitivity of the Ono-Hofmann detection scheme in the presence of a local oscillator of infinite power, as given in Eq.~(\ref{ono_sens}), and the total input photon number $n_{\rm in}$. The presence of contours of value close to one indicates that phase sensitivity with the Ono-Hofmann detection scheme is indeed Heisenberg-limited when the coherent and squeezed vacuum photons alone are counted as photon resource. Figure~\ref{ono_contour_2}, on the other hand, shows the product of phase sensitivity of the Ono-Hofmann scheme in the presence of a local oscillator of finite power, as given in Eq.~(\ref{ono1})), and the square root of the sum of photon numbers in the input and the local oscillator field, $\sqrt{n_t}=\sqrt{n_{\rm in}+n_{\rm lo}}$. It is, however, seen that there exist no contours of value one or less, which indicates that the phase sensitivity is in fact worse than shot noise when the photons in the local oscillator are also accounted for as part of the photon resource. 

At this point, we'd like to mention that the scheme proposed by Plick {\it et al.} in Ref.~\cite{bill} to implement parity detection for interferometry with Gaussian states using balanced homodyning and intensity difference measurement would also suffer from the same drawback when the photons in the local oscillator are counted as resource. Nevertheless, given the fact that these schemes are proposed for applications in the high power regime, the above mentioned drawback is not so consequential. At LIGO, while there is plenty of laser power available, the interferometers are power-limited with the risk of melting the mirrors and beam-splitters. Therefore, the power in the local oscillator in any case cannot be used inside the interferometer. Thus, phase estimation, which is Heisenberg-limited in the photon number inside the interferometer, as provided by the Ono-Hofmann detection scheme or parity measurement inferred through homodyning, would still be appreciated.

\section{Summary}
\label{concl}
We have studied the application of parity detection for phase estimation in Mach-Zehnder interferometry with coherent and squeezed vacuum light. We have shown that parity detection saturates the Quantum Cramer Rao bound of the interferometric scheme and provides Heisenberg-limited phase sensitivity when the coherent and squeezed vacuum light are mixed in equal proportions. Parity can be readily implemented using photon-number-resolving detectors~\cite{migdall} in the low power regime, and possibly using optical nonlinearities and homodying in the high power regime~\cite{nonlinear2, nonlinear, bill}.

We have also presented a brief study of a symmetric-logarithmic-derivative-based detection scheme recently proposed by Ono and Hofmann for the same interferometric scheme in the high power regime~\cite{onoscheme}, with explicit calculations of the signal and phase sensitivity of the scheme. We have shown that this scheme requires a strong local oscillator field in order to provide Heisenberg-limited phase sensitivity. When the local oscillator power is accounted for as part of the photon resource, the phase sensitivity of the scheme is at best short-noise limited---a drawback parity detection implemented using homodyning is also bound to suffer.

\section{Acknowledgments}

KPS acknowledges the Louisiana Board of Regents for funding.

\section{References}

\bibliography{references_1}

\end{document}